# Soft chemistry assisted On-chip Integration of Nanostructured α-quartz-based Piezoelectric Microelectromechanical System

Claire Jolly[1], Andres Gomez[2], David Sánchez-Fuentes[1], Dilek Cakiroglu[1], Raïssa Rathar[1], Nicolas Maurin[1], Ricardo Garcia-Bermejo[1], Benoit Charlot[1], Martí Gich[2], Michael Bahriz[1], Laura Picas[3], and Adrian Carretero-Genevrier[1*]

1. Institut d'Electronique et des Systèmes (IES), CNRS, Université de Montpellier, 860 Rue de Saint Priest 34095 Montpellier, France
2. Institut de Ciència de Materials de Barcelona ICMAB, Consejo Superior de Investigaciones Científicas CSIC, Campus UAB 08193 Bellaterra, Catalonia, Spain
3. Institut de Recherche en Infectiologie de Montpellier (IRIM), CNRS UMR 9004−Université de Montpellier, 34293 Montpellier, France

E-mail: carretero@ies.univ-montp2.fr

**Abstract**

The development of advanced piezoelectric α-quartz MEMS for sensing and precise frequency control applications requires the nanostructuration and on-chip integration on silicon of this material. However, the current quartz manufacturing methods are based on bonding bulk micromachined crystals on silicon, which limits the size, the performance, the integration cost and the scalability of quartz micro devices.

Here, we combine chemical solution deposition, soft-nanoimprint lithography and top-down microfabrication processes to develop the first nanostructured epitaxial (100)α-quartz/(100)Si piezoelectric cantilevers. The coherent Si/quartz interface and film thinness combined with a controlled nanostructuration on silicon–insulator-silicon technology (SOI) substrates provides high force and mass sensitivity while preserving the mechanical quality factor of the microelectromechanical systems. This work proves that biocompatible nanostructured epitaxial piezoelectric α-quartz based MEMS on silicon can be engineered at low cost by combining soft-chemistry and top-down lithographic techniques.

## 1. Introduction

Piezoelectric materials are at the core of many everyday applications thanks to their intrinsic capability to generate electrical charges under and applied mechanical stress or to induce a mechanical deformation from an electrical input. Such properties make them, key elements of the motion detectors and resonators present in many wireless network sensors, which are devices capable of harvesting and transmitting environmental data in an autonomous way. Thus, in this context, piezoelectric materials can find multiple military, security, medical and environmental applications. Today, the monolithic integration of these materials into silicon technology and its nanostructuration to develop alternative cost-effective processes with superior performances are among the central points in current technology[1,2]. Generally, PZT, ZnO and AlN piezoelectric materials are integrated in the form of thin films on Si substrates for chip fabrication. In contrast, quartz-based devices have been so-far micromachined from bulk crystals. This has the disadvantages of limiting their downscaling to thickness below 10 µm[3,4] and that for most of their applications quartz crystals need to be bonded on Si substrate[5]. These drawbacks represent an important bottleneck for the microelectronics industry, since thinner single crystalline quartz plates are currently highly demanded over bulk structures for faster and higher frequency device operations as well as for their capability of lower detection levels with improved sensitivity[4,6,7]. Despite recent achievement to micropattern bulk crystals by laser interference lithography[8], Faraday cage angled-etching and focused ion beam (FIB)[4,9], the integration of thin patterned films on Si is still a challenge.

α-quartz is widely used as a piezoelectric material due to its outstanding properties[5]. While it cannot applied in energy harvesting for its low piezoelectric coefficient, α-quartz presents an excellent chemical stability and high mechanical quality factor make it one of the best candidates for frequency control devices and acoustic sensor technologies[10]. Recently, we developed a direct bottom-up integration of epitaxial α-quartz into (100) silicon substrates[11,12]. Tailoring the structure of α-quartz films on silicon substrates was achieved by chemical solution deposition (CSD)[13], which allowed to control the texture, density and the thickness of the films[14]. As a result, it is now possible to extend the thickness of the α-quartz films from a few hundreds of nanometers to the micron range, which is 10 to 50 times thinner than those obtained by top-down technologies on bulk crystals. Moreover, optimization of the CSD conditions makes it possible to obtain a continuous crystalline quartz nanoimprinted pattern by a combination of a set of top-down lithography techniques[15]. Thus, this opens the possibility to develop cost-effective nanostructured piezoelectric epitaxial α-quartz based MEMS with enhanced sensing properties.

In the present work, we report the fabrication of the first epitaxial piezoelectric nanostructured (100)α-quartz/(100)Si based cantilever, taking advantage of the complementarity of soft-chemistry and top-down lithographic techniques combined with silicon–insulator–silicon (SOI) technology. We have engineered a coherent quartz/silicon interface with a quartz thicknesses above 1000 nm and a controlled nanostructuration, which enable the fabrication of sensitive on-chip quartz devices capable of measuring, thanks to its piezoelectricity, tiny forces and masses through a variation in the resonance frequency while preserving the mechanical quality factor of the device. We also proved that the nanostructured quartz surface is biocompatible and induces the self-organization of proteins on cellular membranes. As a result, nanostructured quartz/Si MEMS could be exploited for bio-sensing applications, for instance, to quantify large number of cellular processes that are sensitive to surface topography (e.g. cell migration, membrane trafficking and signaling and host-pathogen interactions), as previously reported on high-performance glass substrates[16]. Our results demonstrate the implementation of large scale and cost-effective integration of nanostructured piezoelectric epitaxial quartz films into innovative electromechanical devices which can be used for future sensor applications in electronics, biology and medicine.

2. Results and discussions

**Fabrication steps of nanostructured piezoelectric α-quartz based cantilevers**

Aiming to produce nanostructured quartz microcantilevers by silicon micromachining, we have engineered epitaxial nanostructured piezoelectric α-quartz films on (100) silicon–insulator–(100)silicon (SOI) substrate (see figure S1). The thickness of nanostructured piezoelectric α-quartz films was set to 1200 nm, comprising a 600 nm thick quartz dense layer and 600 nm thick nanopillared top layer with diameter and separation distances of 600 nm and 1 µm, respectively (see figure S1). The SOI substrate used along this work was composed of a 2 µm thick silicon active layer, a 500-nm-thick silicon dioxide intermediate layer and a 675-µm-thick base. The piezoelectric activity of the epitaxial quartz films on SOI was evaluated by two different experimental techniques i.e. (i) the traditional piezoelectric force microscopy (PFM) and (ii) the Direct Piezoelectric Force Microscopy (DPFM), a powerful methodology recently developed by the authors which allows to directly measure the piezoelectric effect in thin films[17,18]. Figure S2 shows that the piezoelectric coefficients obtained from both measurements are similar and comparable to that of quartz bulk materials (i.e. 1.5 and 3.5 pm/V)[19]. The crystallinity and crystal orientation of quartz films were analyzed by 2D X-ray diffraction (2D XRD). Figure S3 shows a long range 2D θ-2θ XRD pattern of a 1200 nm thick patterned quartz film on SOI substrate confirming the (100) out of plane texture of α-quartz with no supplementary peaks from other reflections or polycrystallinity signals. The rocking curve of the (100) quartz reflection presented in the figure S3b shows an mosaicity degree of 1.89° , as previously observed in films[11].

Figure 1 illustrates the general schematic of the nanostructured epitaxial (100) quartz film synthesis on SOI substrates with the soft chemistry process followed by the microfabrication of patterned cantilever using silicon-based micromachining technology. The epitaxial growth and nanostructuration of (100) α-quartz films was performed by annealing nanoimprinted Sr-silica xerogel film deposited on (100) SOI substrate according to the dip coating method previously reported[15]. Soft nanoimprint lithography (NIL) was applied on the last Sr-silica gel layer using a PDMS mold to pattern the film surface just before quartz crystallization (see figure S4). The thickness and density of the quartz film was controlled with the number of deposited layers, withdrawal rate, and humidity during synthesis, as previously described[14]. We determined the optimal fabrication steps to produce nanostructured quartz-based cantilevers with a low cost and easy process on SOI substrate, as shown in Figure 1 and Figure S4. The key mechanistic aspects ensuring the preservation of the crystal quality, piezoelectric functionality and mechanical quality factor Q of quartz based cantilevers were (i) the protection of the lateral edges of the patterned epitaxial quartz layer to avoid HF acid infiltration during the release process of the cantilever (see figure S4h and experimental part), (ii) the use of a buffer HF solution (BHF) for a gentle release of quartz based chips and (iii) the increased thickness of the negative resist permitting longer wet etch times. As a result, quartz-based cantilevers present a coherent quartz/silicon interface with uniform epitaxial crystallinity and electromechanical properties, as further described in the following sections.

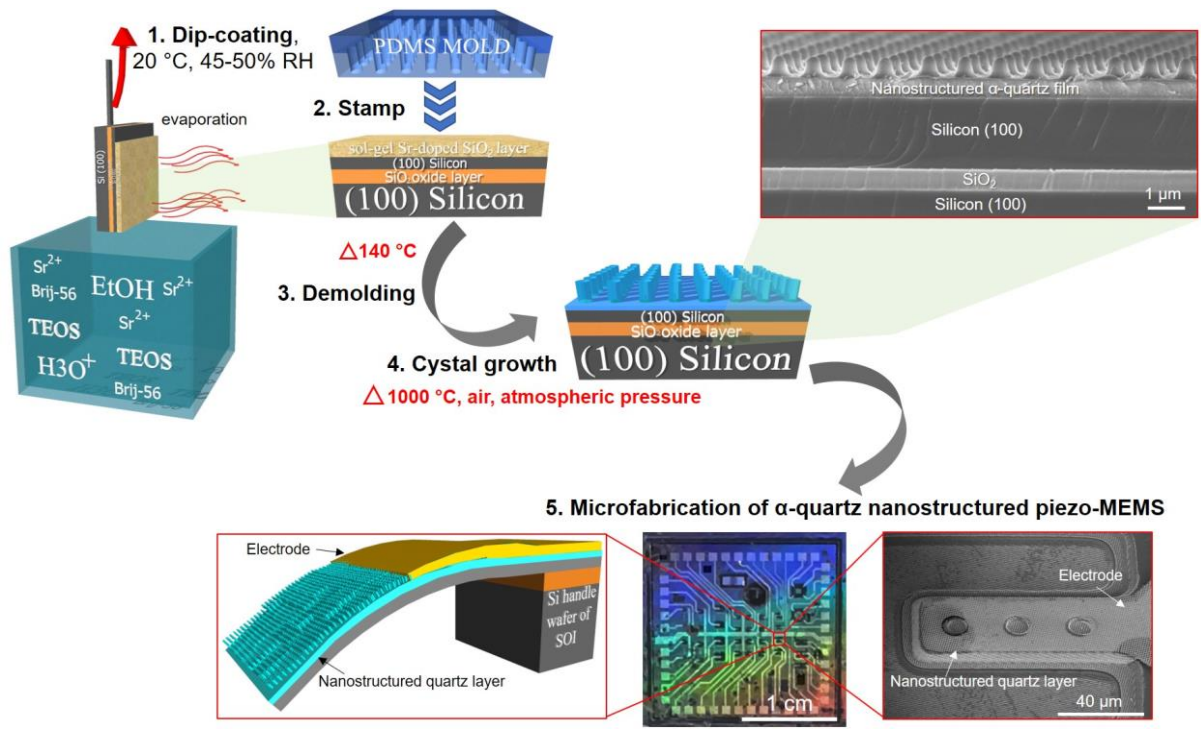

**Figure 1.** General schematics of the synthesis and microfabrication of patterned quartz cantilever. Multilayer deposition of Sr-silica solution on SOI substrate by dip coating is followed by patterning of the film with NIL process. Annealing of the sample at 1000°C enables the crystallization of nanostructured quartz film. Finally, a nanostructured quartz cantilever is fabricated. Notice that colored scattered optical image corresponds to the nanostructuration pattern on the surface of the device that is associated with the diffraction of natural light with the quartz nanopillars.

## Structural and electromechanical characterization of nanostructured piezoelectric α-quartz based cantilevers

To confirm the crystallinity of the quartz layer after MEMS fabrication, 2D X-ray micro diffraction on a single cantilever was performed as seen in Figure 2a. Figure 2b shows a tilted FEG-SEM image of a nanostructured quartz-based cantilever where the different crystalline layer observed in the 2D X-ray diffractogram are highlighted (i.e. (100) oriented epitaxial quartz layer, (100) silicon substrate and the conformal polycrystalline 150 nm thick Pt electrode layer) (see also figure S5). The 2D XRD and rocking curve analysis revealed the same (100) quartz out of plane texture and FWHM value that the films had before the microfabrication process (Figure 2g, 2i). Figures 2c-f display a compositional analysis conducted by energy dispersive x-ray spectroscopy (EDX) elemental mapping. This analysis shows that the cantilever contains a homogeneous composition of Si, O and Pt, which correspond to the piezoelectric quartz layer and the top-electrode coating, whereas silicon is detected in the substrate as a unique element. Therefore, these results confirm the compatibility between MEMS fabrication process and material synthesis. The mechanical response of the quartz based piezoelectric cantilevers was detected by using a Laser Doppler Vibrometer (LDV) equipped with a laser beam, a photodetector and frequency generator (see Figure 2j and experimental section for more information). Among the different cantilever's dimensions included in the fabricated chip (see Figure 1), the nanostructured cantilever analyzed in Figure 2j exhibited a rectangular geometry (40 µm wide and 100 µm long) and was fabricated with a 1200 nm thick patterned quartz layer. We measured a resonance frequency at 267 kHz (comparable to similar commercial tip less cantilevers, see Figure S6) and the estimated quality factor, Q, of the whole mechanical structure was Q ~ 398 under low vacuum conditions. This Q value of nanostructured quartz-based cantilevers is of the same order of magnitude as the mechanical Q of similar and standard non-piezoelectric silicon cantilevers which typically range from 50 to 1000, but can in some cases be as high as 10000[20]. As a result, the epitaxial growth of (100) quartz layer on silicon might not affect the final mechanical Q value of the MEMS structure, probably due to the coherent quartz/silicon interface. Notice that this Q value depends on the external and internal damping phenomena i.e. the cantilever geometry and the damping caused by viscous and other forces coming from the surrending media[21]. Therefore, here we demonstrate the first, to our knowledge, cost-efficient on-chip integration of piezoelectric nanostructured epitaxial quartz-based MEMS. A comprehensive explanation of the methodology and reasoning employed for calculating the quality factor of the nanostructured epitaxial quartz-based cantilevers can be found in the supporting information. We observed a linear dependence of cantilever displacement on the applied AC voltage. The inset of Figure 2f points out the linearity between the vibration amplitude and the actuation voltage at different AC voltages ranging from 1V to 10V with 1V increments. This shows, that these devices can be activated through the converse piezoelectric effect coming from the epitaxial quartz layer.

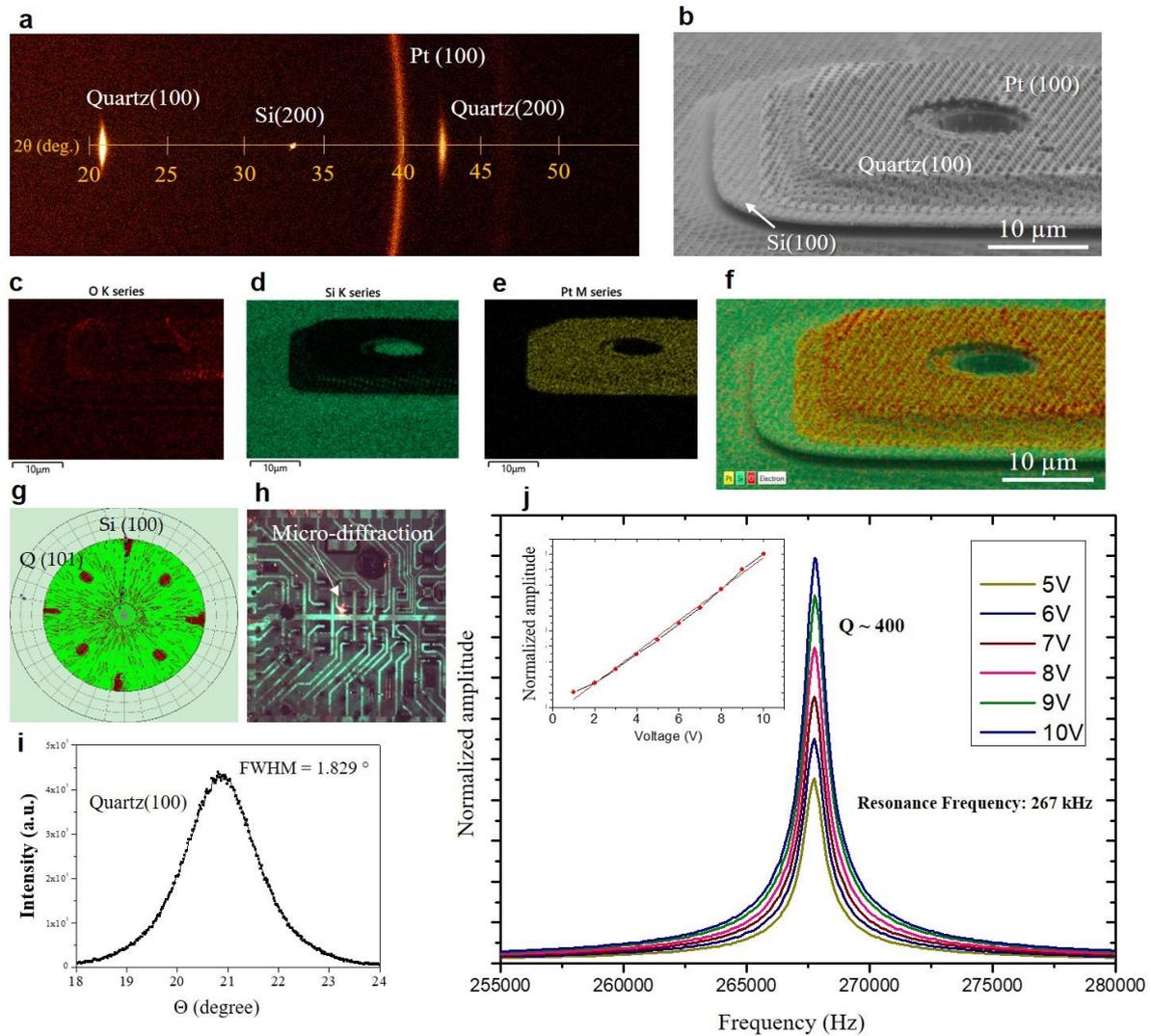

**Figure 2**. (a) Crystallinity of the different materials present in the cantilever. Platinum used for contacts and (100) oriented quartz and silicon substrate are shown in the 2D-XRD diffractogram. (b) Tilted FEG-SEM image of nanostructured cantilevers showing the different layers of used material. (c,d,e and f) Merch of tilted FEG-SEM image and chemical mapping of O, Si and Pt elements of a single quartz/Si cantilever. (g) 2D pole figure of nanostructured α-quartz(100)/Si(100) cantilever. (h) Optical image of the whole chip during microdiffraction measurements at the zone pointed by a laser beam. Notice that the green colour in the optical image corresponds to the diffraction of the natural light produced by the interaction of light and the quartz nanopillar which act as a photonic crystal (i) Rocking curve of the quartz/Si cantilever showing a mosaicity value of 1.829° of the (100) quartz reflection. (j) Electromechanical characterization by noncontact vibrometry measurements under vacuum of a quartz-based cantilever 40 µm wide and 100 µm long composed of a 1200 nm thick patterned quartz layer. The nanopillars diameter and separation distance are 600 nm and 1 µm, respectively and the thickness of the Si device layer is 2 µm thick. The inset image shows the linear dependence of the cantilever amplitude on the applied AC voltage.

## Electromechanical characterization of nanostructured α-quartz cantilevers by AFM

The electromechanical properties of nanostructured quartz cantilevers were also studied under air condition by using atomic force microscopy (AFM) contact measurements. Figure 3a describes the working principle of AFM measurements, where an AC drive output of a Lock-in Amplifier (LIA) is fed to the top and bottom electrodes of the cantilever, while the vibration is recorded with the optical beam deflection system of the AFM. Figure 3b and 3c show the bulk oscillations of the AFM tip before and after routing the signals to the LIA's input i.e 0 V and 10 V, respectively. These results show the active electromechanical response of the quartz based microcantilevers, as already observed in the above described non-contact vibrometry measurements. Moreover, nanometric displacements as high as 12 nm under 10 V were observed confirming that even under air conditions there is a robust mechanical coupling between the Si and the (100) epitaxial quartz layer actuated through the piezoelectric effect (see figure 3d). This feature indicates that the quartz-based cantilever would not be much affected by the air damping phenomena and also by the clamping coming from the AFM tip that remains in contact on the surface of the quartz layer during the measurements. The same linearity between the vibration displacement and the actuation voltage was observed, therefore proving that these devices can be activated through the converse piezoelectric effect (see figure 3e). We analyzed the resonance of the system (i.e. AFM tip and microcantilever) at different positions of the quartz based microcantilever and as expected, the tip vibration amplitude increases closer to the end of the cantilever and disappears when the AFM tip is positioned out of the quartz-based resonator (see figure 3f). Importantly, we have experimentally tested the mass and force resolution of nanostructured quartz-based cantilevers by applying different forces in the µN range with the AFM tip and recording in-situ the resonance frequency evolution (see figure 3g). Considering that Q higher than 100 allows for a frequency resolution below 10 Hz, we obtained a sensitivity of 1 Hz/1µN, which results in a mass detection sensitivity of 100 ng/Hz. As a result, one could use commercial frequency modulators widely used in AFM microscopy, which display µHz resolution, to resolve masses in the pg scale (see figure 3g).

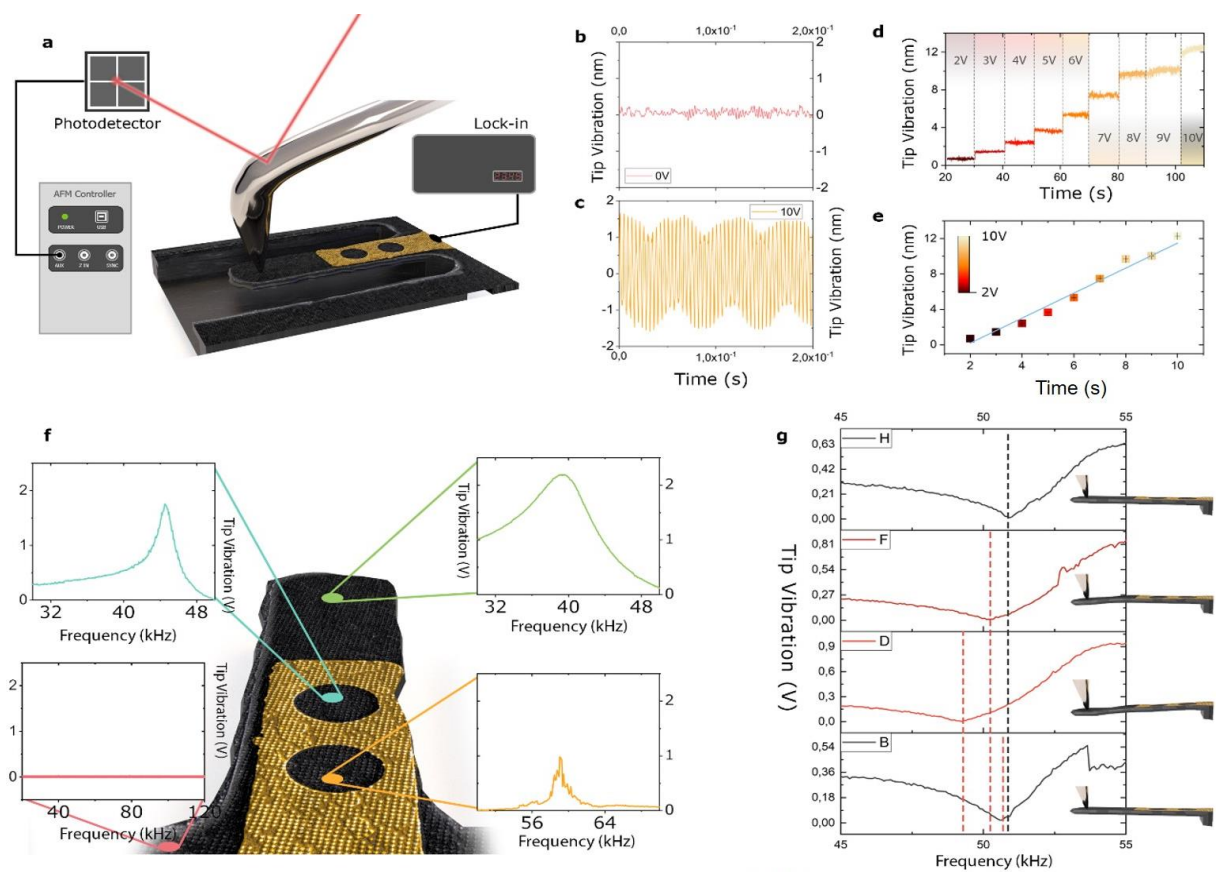

**Figure 3.** Characterization of quartz cantilevers using an Atomic Force Microscope. **a.** The setup of the measurements in which the AC drive output of a Lock-in Amplifier is fed to the top (yellow zone) and bottom electrodes of the sample, while the vibration is recorded with the Optical Beam Deflection System of the AFM. **b** and **c** bulk oscillations of the AFM tip before routing the signals to the LIA's input and after input 10 V signal, respectively. **d**, LIA's amplitude vs Time for different applied voltage amplitudes (from 2 to 10 VAC). **f**, resonance curves obtained from the combination of sample and tip spring constant in different spots of the cantilever. **g**, Evolution of the resonance curve as a function of the applied load on the tip.

Because quartz-based devices are highly used for chemical and bio-sensing applications, such as quartz crystal microbalance (QCM)-based setups[7], we determined the biocompatibility of nanostructured α-quartz thin films engineered by chemical solution deposition, as detailed above (figure 4). We confirmed that human epithelial cells, such as HT1080 cells, adhere and can be cultured on patterned epitaxial α-quartz thin films as shown by scanning electron microscopy (see figure 4a). Next, we compared the growth and proliferation of cells seeded on conventional glass coverslips typically used in biology studies with (100)quartz/(100)Si under cell culture conditions (37ºC and 5% $CO_2$). Whereas (100)Si substrate prevented proper adhesion of epithelial cells, we found that patterned (100)quarz/(100)Si and $SiO_2$ display equivalent biocompatibility, at least up to 84h of under culture conditions, as shown by the quantification of the number of cells per field (N = 6) displayed in figure 4b. Importantly, we found that nanostructured α-quartz thin films can induce the self-organization of the epidermal growth factor receptor (EGFR) on cellular membranes (figure 4c), as confirmed by the 1.05 µm spacing in the 2D-autocorrelation function which is in agreement with the PDMS mold used in this work and with previous studies using advanced fluorescence microscopy on borosilicate glass[16]. As a result, nanostructured biocompatible piezoelectric quartz-based MEMS could be applied for bio-sensing applications in the near future to quantify the contribution of surface topography on cellular processes such as cell migration, membrane trafficking and signaling and host-pathogen interactions, among others.

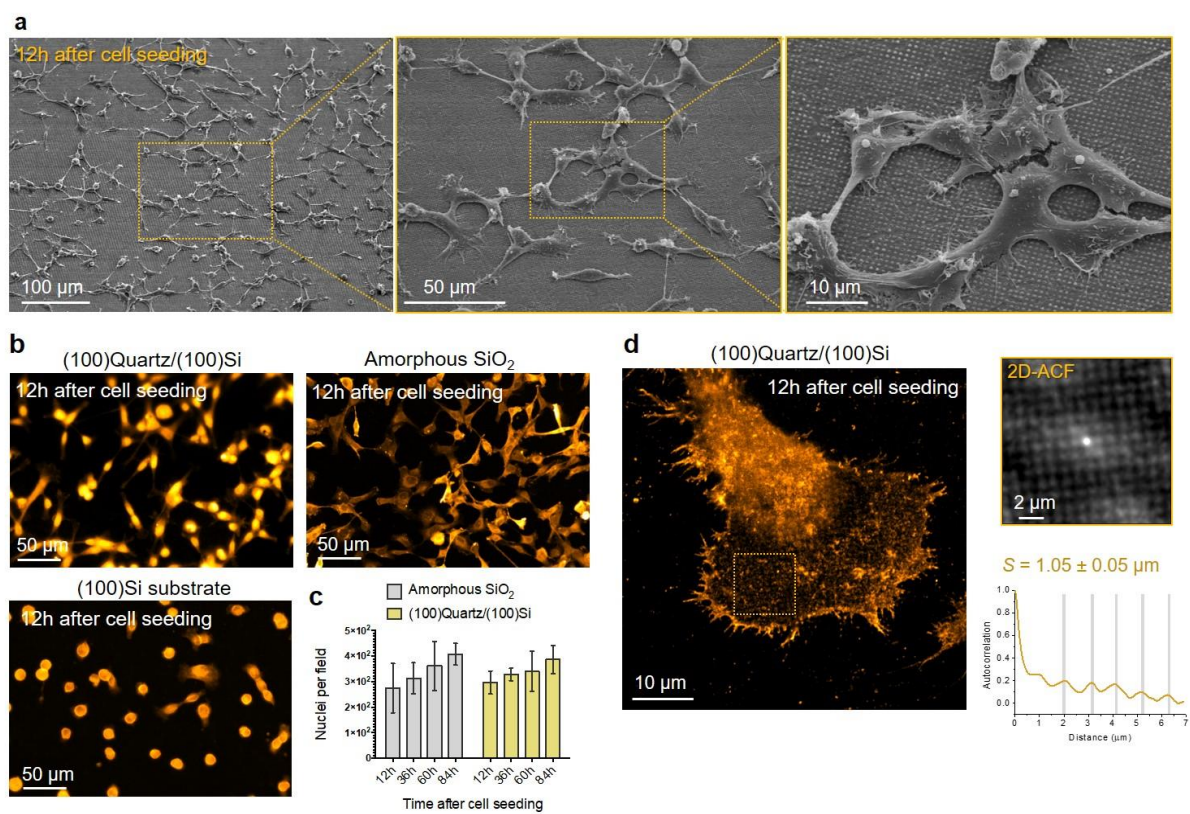

**Figure 4. a)** Representative scanning electron microscopy images displaying HT1080 cells cultured on patterned (100)Quarz/(100)Si during 12h under culture conditions (37ºC and 5% CO2). Scale bar, 10 μm and 5 μm for the magnified images. **b)** Representative widefield images of HT1080 cells expressing EGFR-GFP to visualize the cellular membrane. The same amount of cells was seeded on three different substrates: Amorphous $SiO_2$ (which is the conventional substrate used in microscopy studies), (100)Si substrate and patterned (100)Quarz/(100)Si. Cells were cultured during 12h before imaging. **c)** Quantification of the number of nuclei per field of image at different times (i.e. 12h, 36h, 63h and 84h) after seeding HT1080 cells expressing EGFR-GFP on amorphous $SiO_2$ (gray) and patterned (100)Quarz/(100)Si (yellow). Error bars represent s.d. **d)** Maximum intensity projection of a representative airyscan image of HT1080 cells expressing EGFR-GFP and seeded on patterned (100)Quarz/(100)Si substrates. 2D autocorrelation function and normalized autocorrelation curve of the EGFR signal corresponding to the orange dashed indicated in the airyscan image. Spacing (*S*) is indicated.

Through this technology, these devices are expected to work also at the intrinsic quartz material frequency, which depends on the thickness of the quartz i.e. around 10 GHz for a 800 nm thick resonator[4]. Operating at such high frequency avoids any damping and viscoelastic phenomena coming from the liquid environment, therefore decreasing the energy losses and improving the sensitivity. Moreover, the quality factor of the quartz material operating at this frequency is expected to be huge (Q> $10^6$)[10] and not affected by the liquid damping. Since we obtained a sensitivity of 100 ng/Hz, we anticipate that quartz-based MEMS could be applied for cell mass detection. Indeed, the total mass of cells is ~ 10 pg[22] and most of the experimental approaches for cell mass detection are typically based on either microfabricated resonators or optical methods[23,24]. Therefore, this new technology might overcome the bottlenecks and importantly, the limited working frequencies and sensitivity of the current quartz transducers made from bulk micromachining or hybrid integration methods, making possible to engineer a first nano-quartz balance with super mass resolution.

## 3. Conclusions

In conclusion, here we have combined a cost-effective and scalable chemical method with soft nanoimprint lithography and silicon micromachining to produce novel nanostructured epitaxial piezoelectric quartz based microcantilevers. Using SOI technology, we have been able to modify the dimensions of piezoelectric quartz based cantilevers while preserving a coherent (100)quartz/(100)silicon crystalline interface. Non-contact vibrometry and contact PFM measurements revealed a nanoscale electromechanical motion in air and low-vacuum of quartz-based MEMS activated by the piezoelectric effect. The coherent Quartz/Si interface and the cantilever geometry gives rise to a Q value of 398 under low vacuum conditions, which is of the same order of magnitude as standard non-piezoelectric silicon cantilevers.

From a technological perspective, these novel quartz-based devices open the door to the development of active biocompatible MEMS engineered from lead-free ferroelectric oxide materials. Indeed, we have proved the biocompatibility of nanostructured epitaxial quartz layers, which rendered a cell adhesion and proliferation equivalent to that of conventional high-performance glass coverslips widely used in biology studies. Finally, nanostructured quartz MEMS provide high force and mass sensitivity, making this system a very promising material for many applications in microelectronics.

## 4. Experimental Section

### 4.1. Synthesis of epitaxial nanostructured quartz films

We prepared nanostructured epitaxial (100) α-quartz film on (100) SOI substrate combining two processes: multi deposition of silica solution with dip coating and surface nanostructuration with soft lithography, NIL.

#### 4.1.1. **Solution Preparation**

Firstly, solution A was prepared by adding 0.7 g of Brij-58 into 23.26 g of absolute ethanol, then 1.5 g of HCl (37%), and 4.22 g of tetraethyl orthosilicate (TEOS). Mixture was stirred until the time that it was used. Then, solution B was prepared with 1M $SrCl_2·6H_2O$ and water. Finally, solution C was prepared by adding 275 μL of solution B into 10 mL of solution A and stirred for 10 min. Preparation of Sr-silica films were prepared with deposition of Solution C onto SOI substrate by dip-coating. It is suggested that every 40 minutes new solution C should be prepared to preserve the stability of the deposited solution and also not to compromise the quality of the film. The amount of Sr introduced with solution B is such that, in solution C, the Sr/SiO 2 molar ratio is 0.05 and the final molar composition of TEOS:Brij-58:HCl:EtOH:SrCl 2 is 1:0.3:0.7:25:0.05.

### 4.1.2. Gel Film deposition by dip-coating

First Sr-silica monolayer on SOI substrate was prepared by dip-coating solution C with a 5 mm/s withdrawal rate in a controlled atmosphere. The ND-DC300 dip coater (Nadetech Innovations) equipped with an EBC10 Miniclima device allowed the deposition of Sr-silica at ambient temperature of 25 °C and relative humidity of 40%. After dip-coating, the prepared gel film was consolidated with a thermal treatment for 5 min at 450 °C under the air atmosphere. The monolayer deposition process can be repeated many times depending on the desired film thickness.

### 4.1.3. Soft Nanoimprint Lithography (NIL) Preparation
#### 4.1.3.1. Preparation of Molds

Si masters used as molds in the NIL process were elaborated with different structures and heights using Laser Imprint Lithography (LIL). PDMS (polydimethylsiloxane) reactants (90 wt % RTV141A, 10 wt % RTV141B from BLUESIL) were transferred onto the Si master and dried at 70°C for 1h before unmolding. Molds were degassed under vacuum (10 mbar) for about 20 min.

#### 4.1.3.2. Nanostructuration of film surface with NIL

First, 500 nm of Sr-silica multilayer on SOI was prepared by repeating the monolayer deposition process. Then, one more mono layer was deposited on top of 500 nm of Sr-silica and this last layer was imprinted using PDMS mold without pressure. Peeling of PDMS mold was carried out after the prepared sample was placed in a furnace at 70 °C for 2 min and then at 120 °C for another 2 minutes to produce a column array of 600 nm height. Finally, the nanostructured sample was kept at 450°C for 10 min for consolidation.

### 4.1.4. α-quartz crystallization

Nanostructured sample was introduced into the furnace at 1000 °C at atmospheric pressure and held at this temperature for 300 minutes for crystallization of quartz. Then, the sample was taken out of the furnace after the furnace was cooled down to room temperature.

### 4.2. Quartz Cantilevers microfabrication process

On chip integrated piezoelectric MEMS device which we designed and fabricated is the combination of the α-quartz thin film/nano-pillar and metal electrodes on the SOI substrate purchased from Universal Wafer with a 2 µm thick silicon active layer, a 500-nm-thick silicon dioxide intermediate layer and a 675-µm-thick base. In the design, the width and length of the rectangle shaped cantilevers are 40 µm x 100 µm, 40 µm x 200 µm and 36 µm x 70 µm, respectively.

The α-quartz piezoelectric cantilever was fabricated by microfabrication techniques. The fabrication process mainly included dry etching, top and bottom electrode deposition and wet etching for cantilever releasing. All fabrication steps are shown in Figure S4. Before starting the fabrication process all samples were cleaned in (2:1) $H_2SO_4$:$H_2O_2$ piranha solution for 10 minutes and rinsed with DI water to get rid of the all organic residues on the sample surface.

Fabrication step started with defining the rectangular shape cantilever pattern on the quartz surface using a dry etching process. AZ2070 negative resist was used as an etching mask after EVG 620 UV lithography exposure. Quartz was etched until 2 µm thick silicon by inductively coupled plasma reactive ion etching (ICP-RIE) (model Corial 210 IL) using the $CHF_3/O_2/Ar$ gas mixture at 22 ºC. End point detection (EDP) and also two-point conductivity measurements were used to decide the etching time. The RIE conditions to engrave the sample were the following: 100 W RF power, 200 W LF power, $CHF_3$ 60 sccm/$O_2$ 20 sccm/Ar 10 sccm at 32.5 Torr pressure. The following step was to realize the top and bottom electrode. For this purpose, AZ2020 negative resist was spin coated on the sample surface and patterned using EVG 620 UV lithography for the 20 nm of Cr and 120 nm of Pt metal deposition. The top electrode was directly deposited on the top of the quartz and bottom electrode was deposited on the top of the 2-micrometer thick Si layer on SOI substrate. Fabrication step continued with etching 2 µm thick Si layer around the rectangular shape quartz structure using a dry etching process. In this step, we followed the same etching procedure as in the first step. In the final step, the quartz cantilever suspension from 675-micron Si base was released by a wet etching process. The 500 nm thick $SiO_2$ layer was etched by hydrofluoric acid BOE 7%. AZ2020 was used as an etching mask after 15 minutes of hard baking of the resist at 140°C in order to strengthen the resilience against the acid.

### 4.3. Structural and piezoelectric characterization of patterned quartz films and cantilevers.

The crystalline textures, rocking curve measurements and epitaxial relationship of quartz films and cantilevers were performed on a Bruker D8 Discover diffractometer equipped with a 2D X-ray detector (3 s acquisition each 0.02º in Bragg-Brentano geometry, with a radiation wavelength of 0.154056 nm). The optical images of films and cantilevers were obtained in an Olympus BX51M optical microscope equipped with a Nikon DS-Fi3 camera. The microstructures of the nanostructured films and cantilevers were investigated with a FEG-SEM model Su-70 Hitachi, equipped with an EDX detector X-max 50 $mm^2$ from Oxford instruments. The topography of nanostructured quartz films was studied by AFM in a Park Systems NX-Scanning Probe Microscopy (SPM) unit. Piezoelectric characterization through the direct piezoelectric effect was made by Direct Piezoelectric Force Microscopy in an Agilent 5500LS instrument using a low leakage amplifier (Analog Devices ADA4530) with Platinum solid tips (Rockymountain Nanotechnology RMN-25 PtIr200H). PFM measurements were performed in an Agilent 5500LS using a long-tip shank length tip to diminish electrostatic interaction (RMN 25PtIr300b) while working in the resonant frequency (~ 80 kHz). A Periodically Poled Lithium Niobate sample from Bruker AFM was used as a reference testing platform.

### 4.4. Vibrometry Measurements

The vibration spectra of the fabricated quartz cantilever were evaluated by Laser Doppler Vibrometer (LDV) equipped with laser, photodetector and frequency generator. The vibrometer (OFV-500D, Polytech) was used in the displacement mode with a range of 50nm/V. The frequency generator utilized to actuate the inverse-piezoelectricity of quartz by the cantilever itself was an arbitrary waveform generator Agilent 33250A. The laser was an OFV-534, Polytech. The amplitude measurement of the fabricated quartz cantilever with a dimension of 40 x 100 μm was carried out under vacuum with the specific pressure of $2.6 \times 10^{-2}$ mbar. The vibration spectra were initially obtained over a 100 kHz bandwidth to narrow the frequency window to be able to spot the resonance peak. Once it was found, the same short frequency sweep was performed to measure the vibration amplitude at different AC voltages.

### 4.5. Biocompatibility of substrates and cell culture

HT1080 cells (a gift from N. Arhel and S. Nissole, IRIM, CNRS UMR 9004, Paris, France) were cultured in DMEM GlutaMAX supplemented with 10% fetal calf serum and 100 U·mL$^{-1}$ of penicillin and streptomycin at 37ºC in 5%$CO_2$. Cells were tested negative for mycoplasma.

EGFR-GFP was a gift from Alexander Sorkin (Addgene plasmid #32751). Plasmids were transfected 24h after cell seeding using JetPEI® transfection reagent (Polyplus transfection®) according to the manufacturer's instructions.

Cells seeded on amorphous $SiO_2$, (100)Si or patterned (100)Quarz/(100)Si substrates were fixed in 3.2% PFA in PBS for 10 min at room temperature, then rinsed in PBS twice and incubated for 10 min at room temperature in 1% BSA. Coverslips were then stained for Hoesch to reveal the cell nucleus. Finally coverslips were mounted with a Mowiol® 4-88 (Polysciences, Inc.).

For the quantification of cell growth and proliferation, images were acquired on a Zeiss Axioimager Z2 epifluorescent upright widefield microscope using a 20X Plan Apochromat 0.8 NA (MRI facility, Montpellier). Quantification was performed by using the nuclei staining (Hoesch) under ImageJ. Images for the analysis of protein self-organization at the cell surface were acquired on a Zeiss LSM880 Airyscan confocal microscope (MRI facility, Montpellier). Excitation source used was an Argon laser of 488 nm wavelength. Acquisitions were performed on a 63x/1.4 objective. Multidimensional acquisitions were acquired via an Airyscan detector (32-channel GaAsP photomultiplier tube (PMT) array detector).

**Acknowledgements**
This project has received funding from the European Research Council (ERC) under the European Union's Horizon 2020 research and innovation programme (project SENSiSOFT No.803004). L.P. acknowledges the ATIP-Avenir program for financial support. The authors thank C. André for providing the transfected HT1080 cell line and C. Cazevielle (MRI-COMET, Montpellier) for assistance with biological SEM images. The authors thank D. Montero for performing the FEG-SEM images and Chemical analysis. The FEG-SEM instrumentation was facilitated by the Institut des Matériaux de Paris Centre (IMPC FR2482) and was funded by Sorbonne Université, CNRS and by the C'Nano projects of the Région Ile-de-France. The authors thank Frederic Pichot, David Bourrier and Guilhiem Larrieu for his expertise and advice during the cantilever lithographic processes. The authors also thank Frank Augereau and Eric Rosenkrantz for his advice during vibrometry measurements.